\documentclass[aip,jcp,reprint]{revtex4-1}

\usepackage{amsmath}
\usepackage{amsfonts}
\usepackage{url}
\usepackage{bm}
\usepackage{bbold}
\usepackage{graphics}
\usepackage{paralist}
\newcommand{\bra}{\langle}
\newcommand{\ket}{\rangle}

\begin{document}

\title{Semi-stochastic full configuration interaction quantum Monte Carlo: developments and application}

\author{N. S. Blunt}
\email{nsb37@cam.ac.uk}
\affiliation{University Chemical Laboratory, Lensfield Road, Cambridge, CB2 1EW, U.K.}
\author{Simon D. Smart}
\affiliation{Max Planck Institute for Solid State Research, Heisenbergstra{\ss}e 1, 70569 Stuttgart, Germany}
\author{J. A. F. Kersten}
\affiliation{University Chemical Laboratory, Lensfield Road, Cambridge, CB2 1EW, U.K.}
\author{J. S. Spencer}
\affiliation{Department of Materials, Imperial College London, Exhibition Road, London, SW7 2AZ, U.K.}
\affiliation{Department of Physics, Imperial College London, Exhibition Road, London, SW7 2AZ, U.K.}
\author{George H. Booth}
\affiliation{University Chemical Laboratory, Lensfield Road, Cambridge, CB2 1EW, U.K.}
\affiliation{Department of Physics, King's College London, Strand, London WC2R 2LS, U.K.}
\author{Ali Alavi}
\affiliation{University Chemical Laboratory, Lensfield Road, Cambridge, CB2 1EW, U.K.}
\affiliation{Max Planck Institute for Solid State Research, Heisenbergstra{\ss}e 1, 70569 Stuttgart, Germany}

\begin{abstract}
We expand upon the recent semi-stochastic adaptation to full configuration interaction quantum Monte Carlo (FCIQMC). We present an alternate method for generating the deterministic space without \emph{a priori} knowledge of the wave function and present stochastic efficiencies for a variety of both molecular and lattice systems. The algorithmic details of an efficient semi-stochastic implementation are presented, with particular consideration given to the effect that the adaptation has on parallel performance in FCIQMC. We further demonstrate the benefit for calculation of reduced density matrices in FCIQMC through replica sampling, where the semi-stochastic adaptation seems to have even larger efficiency gains. We then combine these ideas to produce explicitly correlated corrected FCIQMC energies for the Beryllium dimer, for which stochastic errors on the order of wavenumber accuracy are achievable.
\end{abstract}

\maketitle

\section{Introduction}
\label{sec:intro}

Projector quantum Monte Carlo (QMC) methods are important tools in calculating accurate properties of quantum systems.\cite{Foulkes2001,Kalos1962,Booth2009} Such methods involve stochastically applying a projection operator, $\hat{P}$, such that the desired evolution is achieved \emph{on average}. This leads to a stochastic and sparse sampling of the object under consideration, thus reducing the associated memory requirement and often allowing for the study of larger systems than possible with exact, deterministic approaches. While this approach is beneficial in granting access to such systems, the stochastic error decays slowly with simulation time; increasing the efficiency of the sampling therefore allows greater statistical accuracy to be obtained.

A recent projector QMC method, full configuration interaction quantum Monte Carlo (FCIQMC)\cite{Booth2009, Spencer2012, Booth2012}, has been greatly successful in the highly-accurate study of many challenging systems, providing FCI-quality results for systems well out of reach of traditional deterministic FCI approaches. While many traditional projector QMC methods such as diffusion Monte Carlo (DMC) sample the wave function in real space, FCIQMC performs the sampling in a space of discrete basis states. This discrete sampling of the wave function allows efficient annihilation to take place between the walkers, greatly ameliorating the sign problem in many situations\cite{Spencer2012} and removing the need for a fixed node approximation.

A recent article by Petruzielo \emph{et al.}\cite{Petruzielo2012} provided a number of significant advances in FCIQMC, including the introduction of a semi-stochastic approach. In this approach the basis states forming the FCI space are divided into two sets. The projection in the space of states in one set, whose states are deemed to be most significant, is performed exactly. The rest of the projection operator is applied stochastically as in the traditional FCIQMC algorithm. By performing projection in the most important region of the space exactly, the stochastic error on results can be significantly reduced. As the additional memory requirements need not be overwhelmingly large, the approach is still capable of treating systems far beyond those accessible to exact diagonalization, and therefore there were no significant drawbacks which offset this reduction in random error.

In this article we further investigate and apply the semi-stochastic adaptation. In section~\ref{sec:theory} we present a brief overview of the method and in section~\ref{sec:determ_space} suggest a flexible and relatively black box method for partitioning the FCI space. In section~\ref{sec:implementation} we explain how the semi-stochastic adaptation can be implemented in a straightforward and efficient manner in an existing FCIQMC code. Due to the importance of the efficient parallel scaling of FCIQMC we place particular emphasis on this aspect. In section~\ref{sec:results} results are presented. It is demonstrated that the semi-stochastic adaptation need not greatly alter the parallel performance in the current regime of applicability, and we present results for our method of partitioning the FCI space, both in the standard energy estimator and also in the calculation of reduced densities matrices within FCIQMC. Finally, the semi-stochastic approach is used to study the Beryllium dimer, with F12 corrections calculated from reduced density matrices.

\section{Theory}
\label{sec:theory}

The FCIQMC wave function is represented by a collection of walkers which have a weight and a sign and reside on a particular many-electron basis state, which if not specified can be assumed to be a single Slater determinant. The total signed weight of walkers on a state is interpreted as the amplitude of that many-electron basis state in the (unnormalized) FCI wave function expansion. The FCIQMC algorithm consists of repeated application of the projection operator
\begin{equation}
\hat{P} = \mathbb{1} - \Delta \tau (\hat{H} - S \mathbb{1})
\end{equation}
to some initial state, where $\hat{H}$ is the Hamiltonian operator, $\Delta \tau$ is some small time step and $S$ is an energy offset (`shift') applied to the Hamiltonian to control the total walker population. 
With sufficiently small $\Delta \tau$, exact repeated application of $\hat{P}$ will project the initial state to the ground state of $\hat{H}$\cite{Spencer2012}. In FCIQMC, $\hat{P}$ is applied such that the correct projection is only performed on average, thus leading to a stochastic sampling of the ground state wave function.

The projection operator can be expanded in the chosen FCI basis as
\begin{equation}
\hat{P} = \sum_{ij} P_{ij} |i \ket \bra j|.
\label{eq:proj_op}
\end{equation}
In the semi-stochastic adaptation the set of basis states is divided into two sets, $D$ and $S$. We refer to the space spanned by those basis states in $D$ as the \emph{deterministic space}, and refer to the basis states themselves as \emph{deterministic states}. The terms in Eq.~(\ref{eq:proj_op}) can then be divided into two separate operators,
\begin{equation}
\hat{P} = \hat{P}^D + \hat{P}^S,
\end{equation}
where $\hat{P}^D$ refers to the deterministic projection operator,
\begin{equation}
\hat{P}^D = \sum_{i \in D, j \in D} P_{ij} |i \ket \bra j|,
\end{equation}
and $\hat{P}^S$ is the stochastic projection operator containing all other terms. In semi-stochastic FCIQMC, $\hat{P}^D$ is applied exactly by performing an exact matrix-vector multiplication, while $\hat{P}^S$ is applied using the stochastic FCIQMC spawning steps as usual.\cite{fn1}

In order to perform an exact projection in the deterministic space, the walker weights must be allowed to be non-integers. This differs from most previous descriptions of the FCIQMC algorithm thus far. To be clear in notation and terminology, we use $C_i$ to refer to the \emph{signed} amplitude on a state, and $N_i$ to refer to the \emph{unsigned} amplitude (and so $|C_i|=N_i$), which we refer to as the \emph{weight} on the state.

A complete iteration of semi-stochastic FCIQMC is performed as follows, where $\hat{T} = -(\hat{H} - S\mathbb{1})$:
\begin{enumerate}
\item
    \textbf{stochastic projection}: Loop over all occupied states. Perform $\chi_i$ spawning attempts from state $| i \ket$, where $\chi_i$ is specified below. For each spawning attempt, choose a random connected state $| j \ket$ with probability $p_{ij}$, where connected means that $H_{ij} = \bra i | H | j \ket \not= 0$ and $i\not=j$. The attempt fails if both $| i \ket$ and $| j \ket$ belong to $D$,  otherwise a new walker on state $| j \ket$ is created with weight and sign given by $T_{ji} C_i \Delta \tau/p_{ij}$.
\item
    \textbf{deterministic projection}: New walkers are created on states in $D$ with weights and signs equal to $\Delta \tau \boldsymbol{T}^D \boldsymbol{C}^D$, where $\boldsymbol{C}^D$ is the vector of amplitudes currently on states in $D$.
\item
    \textbf{death/cloning}: Loop over all occupied states in $S$. For each state create a spawned walker with weight and sign given by $T_{ii} C_i \Delta \tau$.
\item
    \textbf{annihilation}: Combine all newly spawned walkers with walkers previously in the simulation by summing together the amplitudes of all walkers on the same state.
\end{enumerate}
$\chi_i$ is chosen probabilistically such that its expected value obeys $E[\chi_i]=N_i$. Although other approaches have been used\cite{Petruzielo2012}, in this work we set
\begin{align}
\chi_i &= \lceil N_i \rceil \; \; \textrm{with probability} \; N_i - \lfloor N_i \rfloor, \\
       &= \lfloor N_i \rfloor \; \; \textrm{otherwise},
\end{align}
where $\lceil N_i \rceil$ denotes rounding up and $\lfloor N_i \rfloor$ denotes rounding down. If integer weights are used then this reduces to $\chi_i=N_i$, as used in previous work\cite{Booth2009}.

In order to reduce the memory demands of having a large number of states occupied with a low weight, a minimum occupation threshold, $N_{occ}$, is defined. After all annihilation has occurred, any walkers with a weight less than $N_{occ}$ are rounded up to $N_{occ}$ with probability $N_i/N_{occ}$ or otherwise down to 0. In practice, we always choose $N_{occ} = 1$. The occupation threshold is not applied to deterministic states so that the deterministic projection is applied exactly.

We further use a modification to the initiator adaptation to FCIQMC\cite{Cleland2010, Cleland2011} by allowing all successful spawning events both from and to the deterministic space to survive. This effectively forces all deterministic states to be initiators, which is sensible since these states should selected by their importance (i.e. weight). In the scheme used by Petruzielo \emph{et al.}, the initiator threshold was allowed to vary based on the number of steps since a walker last visited the deterministic space, and so our approach is different (although deterministic states are always initiators in both schemes). In Supplemental Material\cite{fn2} we show that our scheme achieves the same qualitative behavior for the Hubbard model as demonstrated in Ref.~(\onlinecite{Petruzielo2012}). We have not performed a comprehensive study of the effect of semi-stochastic on the initiator error. However, we tend to find that when the number of walkers, $N_w$, is much larger than the deterministic space size, the use of semi-stochastic makes little difference. This is expected because the two approximations should be identical in the limit $N_w$/$|D| \gg 1$. We also note that it is not essential to use both the initiator and semi-stochastic adaptations together; the benefits from both extensions are largely independent of each other. However, all results presented in this article do use the initiator adaptation.

Using non-integers weights can have a significant memory impact compared to integer weights due to the large number of additional spawned walkers, which also increases time demands due to expensive extra processing and communication steps. We therefore apply an unbiased procedure to stochastically remove newly-spawned walkers with very small weights, similar to that above. Following the notation of Overy \emph{et al.}\cite{Overy2014}, we use a spawning cutoff, $\kappa$, where  $\kappa = 0.01$ unless stated otherwise.  A spawning of weight $N_j < \kappa$ is rounded up to $\kappa$ with probability $N_j/\kappa$ or otherwise down to 0. Spawned walkers with weights greater than $\kappa$ are left unaltered.

\section{Choosing the deterministic space}
\label{sec:determ_space}

The key to reducing stochastic error within the semi-stochastic approach is to choose $D$ such that most of the weight of the true FCI wave function is in this space. For a given number of basis states in the deterministic space, $|D|$, it is expected that the best possible deterministic space (the one which reduces noise the most) is obtained by choosing the $|D|$ most highly weighted basis states in the exact expansion of the ground-state wave function. Achieving this optimal space requires knowledge of the exact wave function and so is not feasible in general.

A sensible choice for $D$ in many systems would be a truncation of the FCI space by number of excitation operators applied to an initial dominant configuration (generally the Hartree-Fock state), giving the truncated `CI' expansion, or alternatively a complete active space (CAS) truncation. These are generally regarded as being effective at describing situations where dynamical and static correlation, respectively, are important. We have found from experience that such spaces are useful and lead to a large reduction in stochastic noise.  This leads to the question: can one find a better deterministic space, at least in common cases?

Petruzielo \emph{et al.}\cite{Petruzielo2012} describe an iterative method for choosing the deterministic space. 
First the space connected to the states chosen in the previous iteration is generated and the ground state of the Hamiltonian in this subspace is calculated. 
The most significant basis states in the ground-state expansion are kept (according to a criterion on the amplitude of coefficients).  The initial space contains (e.g.) the Hartree--Fock determinant.  This process is repeated for some number of iterations. This approach was shown to give much greater improvements than by simply using the space connected to the Hartree--Fock state, even with a reduced size for $D$, as it can contain the chemically-relevant basis states\cite{Petruzielo2012}.

In this work we present and use a new method of generating the deterministic space. Inspired by the spirit of FCIQMC, we allow the deterministic space to emerge from the calculation itself: we simply perform a fully-stochastic FCIQMC calculation (or a semi-stochastic calculation with a simple deterministic space, such as a CISD space) until a coarse representation of the ground state is deemed to have been reached, and then choose the most populated basis states in the FCIQMC wave function to form $D$. Because the semi-stochastic adaptation does not significantly change the rate of convergence, and statistics are not accumulated until the ground state is reached anyway (where the semi-stochastic adaptation is of more benefit), this requires no extra computational effort. This approach has the benefit that it does not require performing an exact ground-state diagonalization within a (potentially large) subspace, which can be very expensive. The only parameter is the desired deterministic space size and it is therefore also a relatively black box approach.

Although the FCIQMC wave function is only a stochastic snapshot of the true ground state, the most significant basis states in the expansion will tend to remain highly occupied throughout the simulation with weights fluctuating about their exact values. The FCIQMC simulation naturally picks out chemically-important determinants, even when deep in the Hilbert space (on quadruple, sextuple and higher excitation levels), and so our procedure can select close-to-optimal deterministic spaces in a very inexpensive manner. For very large deterministic spaces, states with the smallest occupation weights may be included in the space. In this case there is some redundancy in how $D$ is chosen, and the choice of $D$ will probably not be optimized fully, although we still find this approach to work very well. It is simple to include a cutoff to avoid this if desired, although we do not do so in the calculations presented here. With our approach we avoid the need for diagonalization steps, which would become unfeasible for large $D$ and as the connectivity of the Hamiltonian grows. For instance, in some previous applications of FCIQMC the number of connections to the Hartree-Fock has been $\mathcal{O}[10^5-10^6]$\cite{Shepherd2012_2,Thomas2015}.

\section{Implementation details}
\label{sec:implementation}

An in-depth description of our FCIQMC implementation is given in Ref.~(\onlinecite{Booth2014}); we present here only the additions to that algorithm required for the semi-stochastic implementation. Our implementation of FCIQMC is parallelized using MPI.  A given basis state is assigned to a particular MPI process, which performs all spawning from that basis state. Deterministic and stochastic states are treated equally in this respect.

Iterative diagonalization methods, such as the Davidson or Lanczos methods, typically require at most a few tens of iterations.  Given the desire to treat as large a system as possible and the memory cost of storing even a compressed form of $\boldsymbol{H}$, many deterministic subspace methods use the direct CI approach of constructing $\boldsymbol{H} \boldsymbol{v}$ as needed\cite{Olsen1988}.
In contrast, FCIQMC calculations regularly require on the order of $10^5-10^6$ iterations. Thus, it is of critical importance that each multiplication by the (comparatively small) deterministic Hamiltonian is very fast. Storing the Hamiltonian, which speeds up this multiplication considerably, is therefore worthwhile and feasible.
Because FCIQMC scales well to large numbers of processors, a large amount of distributed memory is typically available and one is usually far more time-limited than memory-limited. As such, we have not found this memory requirement to become an issue.

The deterministic Hamiltonian is stored in a sparse matrix format and split across processes so that, if $|i \ket$ belongs to an MPI process, then all non-zero elements $\bra i | H | j \ket, |j\ket \in D$ (i.e. the entire compressed row) are also stored in memory on that process. In contrast the walker amplitude for basis state $|i \ket$ is only stored in memory for the process to which $|i \ket$ belongs.  We gather the amplitudes of the deterministic basis states via an \url{MPI_AllGatherV} call and perform the deterministic projection via a sparse matrix multiplication on each MPI process.  Our implementation therefore requires one additional parallel communication per iteration compared to the standard FCIQMC algorithm.

For the most part, deterministic states are treated in the same manner as non-deterministic states. Because the stochastic and deterministic spaces are coupled, stochastic spawning attempts must still be performed from $D$. However, because the deterministic-to-deterministic projection is performed exactly, such stochastic attempts should not create new walkers inside $D$.  For simple deterministic spaces, such as CI and CAS spaces, it is possible to create excitation generators which never create deterministic-to-deterministic spawnings. This is not feasible for more general spaces, such as in the schemes outlined in the previous section.  Instead, we remove any walkers stochastically spawned from a state in $D$ to another state in $D$.  This check is efficiently performed by using a hash table (similar to that used for annihilation in FCIQMC\cite{Booth2014}) of the deterministic space, such that the test of whether the basis state is in $D$ can be performed in $\mathcal{O}[1]$ time.  The extra memory required to store the hash table is usually negligible compared to other memory requirements, such as that of the deterministic Hamiltonian. It is also partially compensated by the fact that a smaller number of (stochastic) spawned walkers are accepted, and so memory demands of the spawned list are decreased.

\section{Results}
\label{sec:results}

\subsection{Parallel performance}
\label{sec:results_par}

\begin{figure}
\includegraphics{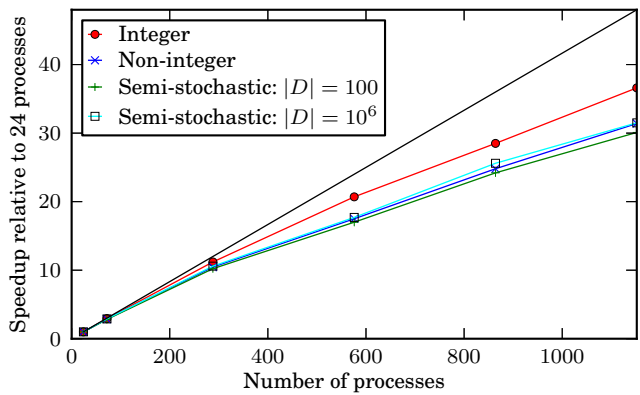}
\caption{Speed-up as the number of the MPI processes is increased from 24 to 1152, for calculations performed on the chromium dimer in an SV basis, with approximately $2\times 10^8$ walkers used. Scaling worsens with the use of non-integer weights, primarily due to exaggerating the poor balancing of work among processes. The semi-stochastic adaptation does not greatly alter the scaling further.}
\label{fig:scaling_cr2}
\end{figure}

Figure~\ref{fig:scaling_cr2} presents the parallel speed-up for the chromium dimer (bond length 1.5\AA, SV basis, CAS (24,30)) from 24 to 1152 cores on ARCHER, a Cray XC30. This system has a Hilbert space size of $\sim \mathcal{O}[10^{14}]$, and approximately $2\times 10^8$ walkers were used in each simulation (sufficient to converge the initiator error to high accuracy).  We consider FCIQMC calculations using both integer weights and non-integer weights, and semi-stochastic FCIQMC using $D=100$ and $D=10^6$. It is apparent that the scaling quality reduces somewhat by using non-integer coefficients. However, there is almost no further decrease in quality when using the semi-stochastic adaptation. In fact, the scaling is slightly improved when using a deterministic space size of $10^6$. The semi-stochastic initialization times, usually negligible compared to the total calculation time, were not included in these results so that the scaling curves did not depend on the number of iterations performed.

By far the largest cause of loss in parallel efficiency in FCIQMC is poor load balancing. The number of basis states and walkers assigned to each process is not precisely constant. Each process therefore takes a different amount of time to complete each iteration. The slowest process acts as a bottleneck for other processes with less work, which must synchronize before parallel communication can be performed. It is found that using non-integer coefficients somewhat exacerbates this issue, leading to the loss of parallel efficiency seen in figure~\ref{fig:scaling_cr2}. This worsening of load balancing with non-integer coefficients is primarily due to the greatly increased number of spawning events that are received, and must be processed, by processes with already-heavy loads. As expected, communication time is also increased by using non-integer weights due to the extra spawning events that must be sent and received, but this time still remains quite small compared to the synchronization time.

\begin{figure}
\includegraphics{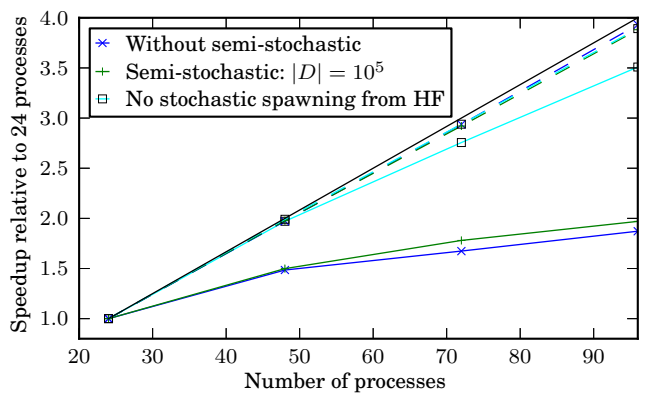}
\caption{Speed-up as the number of the MPI processes is increased from 24 to 96, for calculations performed on the 18-site Hubbard model with approximately $5 \times 10^7$ walkers. Solid lines show results at $U/t=1$ and dashed lines show $U/t=8$. At $U/t=1$ scaling is very poor due the extremely large number of walkers on the Hartree--Fock state. The scaling is improved slightly with the use of semi-stochastic, as far fewer spawnings from the Hartree--Fock state survive. If all connections to the Hartree--Fock are included in the deterministic space then stochastic spawning from the Hartree--Fock does not need to be performed. With this change the load balancing is greatly improved. At $U/t=8$ the scaling is much better, and little difference is made by the semi-stochastic adaptation.}
\label{fig:scaling_hubbard}
\end{figure}

Figure~\ref{fig:scaling_hubbard} shows similar results for the 18-site 2D Hubbard model, with solid (dashed) lines representing $U/t=1$ ($U/t=8$). All calculations used approximately $5 \times 10^7$ walkers. 
The load balancing for $U/t=8$, which has a highly delocalized wave function in the Bloch basis, is very good. Excellent (essentially identical) parallel scaling is seen in all cases.
The $U/t=1$ system, however, is heavily dominated by the Hartree--Fock state. Therefore, the process to which this state belongs takes much longer to complete each iteration than other processes. This problem is exacerbated as the process count increases, and the parallel performance is very poor. It is found that using the semi-stochastic adaptation slightly improves this performance, as the Hartree--Fock process has far fewer successful spawning events to perform calculations on, due to many deterministic-to-deterministic stochastic spawning events being canceled. Despite this, the most expensive steps are the spawning attempts, which are still performed, and so the load balancing is still poor.

However, if all states connected to the Hartree--Fock through a single application of $\hat{H}$ are included in the deterministic space, then \emph{all} stochastic spawnings from the Hartree--Fock state will be canceled. As such there is no need to attempt any spawning from the Hartree--Fock state. When this change is made, the parallel performance improves dramatically. This suggests that using deterministic approaches to treat the most heavily-weighted states may be a very effective way to improve the parallel performance of FCIQMC. This approach has not been used for any further results in this article, but will be an area of research going forward.

\subsection{Hartree--Fock energy estimator}
\label{sec:results_hf}

We first present efficiency increases for the standard Hartree--Fock-based projected energy estimator:
\begin{equation}
E_0 = \frac{\bra D_{\textrm{HF}} | H | \Psi \ket}{\bra D_{\textrm{HF}} | \Psi \ket},
\label{eq:hf_estimator}
\end{equation}
where $|D_{\textrm{HF}} \ket$ is the Hartree--Fock state and $|\Psi \ket$ is wave function represented by the FCIQMC walkers. The efficiency measure that we consider is the same one used by Petruzielo \emph{et al.}\cite{Petruzielo2012},
\begin{equation}
\epsilon = \frac{1}{\sigma_{\mu}^2 \times T},
\end{equation}
where $T$ is the total simulation time (excluding initialization time\cite{fn3}) and $\sigma_{\mu}$ is the final energy error estimate, obtained by averaging multiple estimates taken from throughout the simulation. This is a sensible efficiency measure because the error of such an average can be estimated via
\begin{equation}
\sigma_{\mu} = \frac{\sigma}{\sqrt{N}},
\end{equation}
where $N$ is the number of \emph{uncorrelated} estimates contributing to a mean estimate (possibly taken from a blocking analysis\cite{Flyvbjerg1989}), and $\sigma$ is the true standard deviation of the distribution from which each estimate is taken, which is constant at equilibrium. $N$ scales linearly with $T$, and therefore this efficiency measure is appropriate.

Caution should be taken in interpreting the relative efficiency of two simulations, however. An increase of efficiency of $X$ with semi-stochastic does not necessarily mean that a particular value of $\sigma_{\mu}^2$ can always be obtained with $X$ times less simulation time by using the semi-stochastic adaptation.  For the efficiency results presented here, all error estimates are calculated through a blocking analysis\cite{Flyvbjerg1989} in order to take account of the serially correlated nature of the FCIQMC wave function between subsequent iterations.  Such an analysis typically requires a large number of iterations (depending on the system and time step used) in order to obtain an accurate error estimate.  Because using semi-stochastic does not seem to reduce the auto-correlation time, a similar number of iterations must be performed.  Thus after a sufficient number of iterations, the error obtained even without the use of the semi-stochastic adaptation may be suitably small.  In such cases, the overhead of performing the deterministic projection is not particularly advantageous.  In most cases, however, more than a small number of independent blocks of data are required, and so using the semi-stochastic approach is highly beneficial.

We present results for a variety of systems to demonstrate the general applicability of semi-stochastic and our method for generating the deterministic space. All calculations were performed with $10^6$ walkers unless stated otherwise, and the deterministic space was generated using the new scheme outlined in section III. The relative efficiencies presented are relative to otherwise identical simulations that do not use the semi-stochastic adaptation (but \emph{do} use non-integer coefficients, with the same value of spawning cutoff, $\kappa$). All calculations for each system were performed on the same machine and number of CPU cores (between 24 and 96). Simulations were typically performed for between $10^5$ and $10^6$ iterations.

In each case the time step was set just small enough to avoid the possibility of bloom events, where more walkers than the initiator threshold are created from one spawning attempt. Such events are undesirable because they instantly become initiators, increasing the associated initiator error as a consequence.

\begin{figure}
\includegraphics{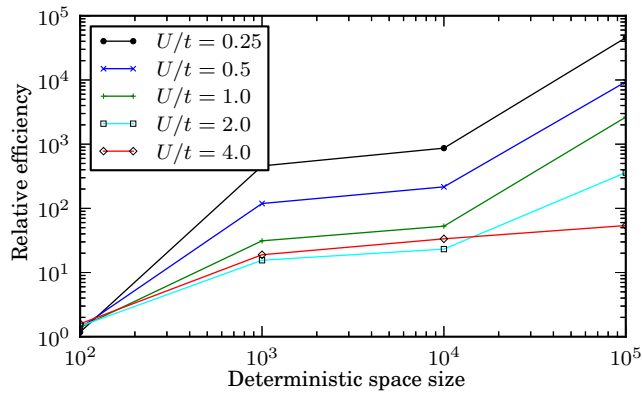}
\caption{The efficiency ($\epsilon_{E_0}$) of semi-stochastic simulations relative to otherwise identical simulations without semi-stochastic, for an 18-site Hubbard model. For small $U/t$ values, where the wave function is dominated by a small number of basis states, the semi-stochastic adaptation helps greatly, leading to an efficiency increase of over $4.5\times10^4$ at $U/t$=0.25 and $|D|=10^5$. For larger $U/t$ values the benefit becomes less significant.}
\label{fig:hub_hf}
\end{figure}

Figure~\ref{fig:hub_hf} shows the efficiency of semi-stochastic FCIQMC for the 18-site 2D Hubbard model at a variety of coupling strengths, from $U/t=0.25$ to $U/t=4$, and for deterministic space sizes ranging from $10^2$ to $10^5$. Significant increases in efficiency are observed in all cases, with the most significant gains occurring at small $U/t$. This is expected: at small coupling strengths the wave function is dominated by a small number of significant determinants which are treated exactly by the deterministic space, whereas at large $U/t$ the equivalent deterministic space will be significantly less occupied.

Table~\ref{tab:hf_efficiency} contains results for two molecular systems, N$_2$ in a cc-pVDZ basis (with 4 core electrons uncorrelated) and Be$_2$ in a cc-pCVTZ basis, at equilibrium and stretched geometries. Once again significant improvements are observed, even at stretched geometries where the wave function is more multi-reference.

\begin{table}
\begin{center}
{\footnotesize
\begin{tabular}{@{\extracolsep{4pt}}cccccc@{}}
\hline
\hline
& \multicolumn{2}{c}{N$_2$} & \multicolumn{3}{c}{Be$_2$} \\
\cline{2-3} \cline{4-6}
$|D|$ & Equilibrium & Stretched & 2.45\AA & 3.0\AA & 5.0\AA \\
\hline
$10^2$ & 6.0    & 30.6   & 2.5   & 3.9   & 2.6   \\ 
$10^3$ & 45.3   & 127.0  & 8.6   & 13.4  & 11.3  \\ 
$10^4$ & 283.4  & 2793.5 & 50.4  & 90.4  & 103.1 \\ 
$10^5$ & 1550.4 & 4710.4 & 218.3 & 765.4 & 560.5 \\ 
\hline
\hline
\end{tabular}
}
\caption{The efficiency ($\epsilon_{E_0}$) of semi-stochastic simulations relative to an otherwise identical simulation without semi-stochastic. Results are shown for N$_2$ in a cc-pVDZ basis with 4 core electrons uncorrelated, at equilibrium (2.118a$_0$) and stretched (10.4$a_0$) geometries, and also for Be$_2$ in a cc-pCVTZ basis at equilibrium and two stretched geometries. A significant increase in efficiency is found for all geometries, and a monotonic improvement with $|D|$ is always observed.}
\label{tab:hf_efficiency}
\end{center}
\end{table}

\begin{figure}[b]
\includegraphics{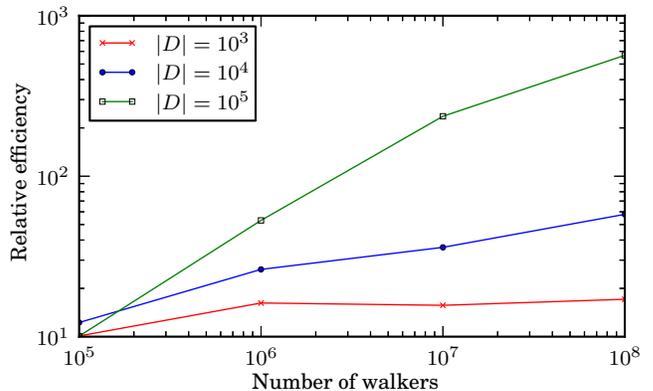}
\caption{The efficiency ($\epsilon_{E_0}$) of semi-stochastic simulations relative to an otherwise identical simulation without semi-stochastic, for the 14-electron homogeneous electron gas with 114 spin orbitals and $r_s=1.0$ a.u., as the walker population is varied. It is found that the benefit of semi-stochastic tends to increase as the walker population increases, contrary to a simplistic intuition that there should be diminishing returns as stochastic error decreases due to the improved stochastic sampling.}
\label{fig:ueg_hf_vary_Nw}
\end{figure}

One might expect that as the number of walkers increases, the improvement gained from semi-stochastic would decrease. This seems reasonable because, as the walker number is increased, the stochastic spawning will approach exact projection. Interestingly, we find the opposite behavior. In figure~\ref{fig:ueg_hf_vary_Nw} the relative efficiency is presented for the homogeneous electron gas\cite{Shepherd2012_1, Shepherd2012_2, Shepherd2012_3} with 14 electrons, 114 plane wave spin orbitals and $r_s = 1.0$ a.u., as the walker population is varied from $10^5$ to $10^8$. For deterministic space sizes of $10^4$ and $10^5$ a significant increase in stochastic efficiency is observed as the walker population is increased. For $|D|=10^3$ the increase is less significant, becoming approximately constant for populations from $10^6$ to $10^8$. These results suggest that the semi-stochastic adaptation will continue to be beneficial for very large systems and walker populations. It is hard to isolate precisely why these benefits increase, but various factors such as the sign problem and subtleties in the impact of the initiator criterion are likely to play a role. These are also expected to be very system-dependent.

\subsection{Reduced density matrix estimators}
\label{sec:results_rdm}

A recent article by Overy \emph{et al.}\cite{Overy2014} introduced a method of unbiased sampling for reduced density matrices in FCIQMC. In this approach a replica sampling\cite{Overy2014, Zhang1993, Hastings2010, Blunt2014} is used, where two independent FCIQMC simulations are performed simultaneously, each starting from a different random number seed. Because these two simulations are statistically independent, it is possible to sample quantities that depend quadratically on the ground-state wave function without introducing a bias.  In particular, the components of the second-order reduced density matrix can be expressed as
\begin{align}
\Gamma_{pq,rs} &= \bra \Psi| a_p^{\dagger} a_q^{\dagger} a_s a_r | \Psi \ket, \\
               &= \sum_{ij} C_i C_j \bra D_i | a_p^{\dagger} a_q^{\dagger} a_s a_r | D_j \ket.
\end{align}
This can be estimated in FCIQMC via
\begin{equation}
\Gamma_{pq,rs} = \sum_{ij} C_i^1 C_j^2 \bra D_i | a_p^{\dagger} a_q^{\dagger} a_s a_r | D_j \ket.
\label{eq:rdm_contribs}
\end{equation}
where $\boldsymbol{C}^1$ and $\boldsymbol{C}^2$ are the walker amplitudes coming from simulations 1 and 2, respectively\cite{fn4}, and $p$, $q$, $r$ and $s$ denote spin orbital labels.

In the implementation used for the results in this article (\url{NECI}\cite{fn5}), diagonal elements ($\Gamma_{pq,pq}$) are calculated exactly for the pair of FCIQMC wave functions used. For non-diagonal elements of $\Gamma_{pq,rs}$, contributions in Eq.~(\ref{eq:rdm_contribs}) including the Hartree--Fock state ($C_{\textrm{HF}}^1 C_j^2 \bra D_{\textrm{HF}} | a_p^{\dagger} a_q^{\dagger} a_s a_r | D_j \ket$) are always included in the estimate, while all other contributions are stochastically sampled alongside the stochastic sampling of the Hamiltonian operator, as described in Ref.~(\onlinecite{Overy2014}). Thus, there are two sources of error contributing to each estimate of the 2-RDM: the random sampling of the ground-state wave function, and also the random sampling of the 2-RDM given these wave functions.

In the semi-stochastic adaptation we modify the estimation of the 2-RDM by also always including all contributions between states in the deterministic space. That is, the contribution
\begin{equation}
C_i^1 C_j^2 \bra D_i | a_p^{\dagger} a_q^{\dagger} a_s a_r | D_j \ket
\end{equation}
is always included if both $|D_i \ket$ and $|D_j \ket$ belong to $D$. This is achieved by storing a further array, roughly the same size as the deterministic Hamiltonian, which stores the excitation levels and coupling parities of pairs of connected states in the deterministic space. This then speeds up the calculation of these contributions, although these quantities could easily be calculated on-the-fly to save memory. Because both $|D_i \ket$ and $|D_j \ket$ belong to $D$, $C_i^1$ and $C_j^2$ should both be large and so semi-stochastic naturally picks out the large contributions to the RDM. Because each process stores all deterministic amplitudes (as a result of the \url{MPI_AllGatherV} call during the deterministic projection step) no extra communication is required.
We therefore emphasize that the semi-stochastic adaptation improves the RDM estimates in two ways, firstly by improving the underlying FCIQMC wave functions and secondly by improving the sampling of the 2-RDM given these wave functions. We note that in our implementation the 1-RDM contribution is calculated from the 2-RDM, and therefore also benefits from the improved calculation of off-diagonal elements of $\Gamma_{pq,rs}$.

\begin{figure}
\includegraphics{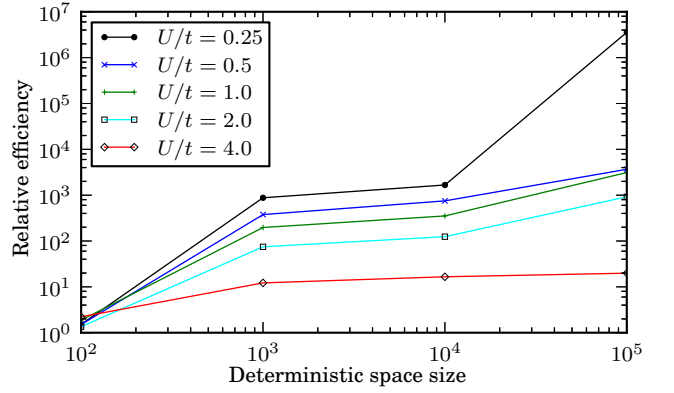}
\caption{The efficiency ($\epsilon_{\bra S^2 \ket}$) of semi-stochastic simulations relative to an otherwise identical simulation without semi-stochastic, for an 18-site Hubbard model. This efficiency measure uses the estimate of $\bra S^2 \ket$ obtained from stochastically-sampled RDMs. The stochastic efficiency is seen to improve in a manner similar to $\epsilon_{E_0}$, although the improvement is even greater.}
\label{fig:hub_rdm}
\end{figure}

\begin{table*}[t]
\begin{center}
{\footnotesize
\begin{tabular}{@{\extracolsep{4pt}}ccccccccc@{}}
\hline
\hline
& \multicolumn{4}{c}{RDM energy estimator} & \multicolumn{4}{c}{RDM $\bra S^2 \ket$ estimator} \\
\cline{2-5} \cline{6-9}
& \multicolumn{1}{c}{N$_2$} & \multicolumn{3}{c}{Be$_2$} & \multicolumn{1}{c}{N$_2$} & \multicolumn{3}{c}{Be$_2$} \\
\cline{2-2} \cline{3-5} \cline{6-6} \cline{7-9}
$|D|$ & Equilibrium & 2.45\AA & 3.0\AA & 5.0\AA & Equilibrium & 2.45\AA & 3.0\AA & 5.0\AA \\
\hline
$10^2$ & 1.82   & 3.57   & 2.03   & 2.55    & 4.23    & 9.20    & 18.57    & 6.89    \\ 
$10^3$ & 46.12  & 6.69   & 11.39  & 12.68   & 28.49   & 411.56  & 524.27   & 89.19   \\ 
$10^4$ & 434.98 & 109.58 & 35.67  & 43.23   & 172.95  & 2363.88 & 8299.70  & 933.13  \\ 
$10^5$ & 855.23 & 231.12 & 234.48 & 1033.02 & 1370.06 & 5877.94 & 11337.21 & 1627.39 \\ 
\hline
\hline
\end{tabular}
}
\caption{Relative efficiencies in the RDM estimates of energy and $\bra S^2 \ket$ for N$_2$ (in a cc-pVDZ basis with 4 electrons uncorrelated, at a separation of 2.118$a_0$) and Be$_2$ in a cc-pCVTZ basis (with all electrons correlated). Large improvements are seen for all systems, geometries and estimators, with a monotonic increase with $|D|$ in each case.}
\label{tab:rdm_efficiency}
\end{center}
\end{table*}

Two separate quantities are considered to study the quality of these RDM estimates. The first is the variational energy estimate
\begin{align}
E_{\textrm{RDM}} &= \bra \Psi | \hat{H} | \Psi \ket \\
        &= \sum_{pq} h_{pq} \gamma_{pq} + \sum_{p>q,r>s} \Gamma_{pq,rs} \bra pq || rs \ket + h_{\textrm{nuc}},
\label{eq:rdm_energy}
\end{align}
where $\gamma_{pq} = \bra \Psi | a_p^{\dagger} a_q | \Psi \ket$ is the 1-RDM. This should be an important quantity in FCIQMC because it is variational, whereas the energy obtained in $i$-FCIQMC from equation~(\ref{eq:hf_estimator}) is not. 
The second quantity considered is the expectation value of $\hat{S}^2$, which for spin-$1/2$ particles can be calculated as
\begin{multline}
\bra \Psi | \hat{S}^2 | \Psi \ket = \sum_{IJ}[ \frac{1}{4} \Gamma_{I\alpha J\alpha, I\alpha J\alpha} + \frac{1}{4} \Gamma_{I\beta J\beta, I\beta J\beta} \\ - \frac{1}{2} \Gamma_{I\alpha J\beta, I\alpha J\beta} - \Gamma_{I\alpha J\beta, J\alpha I\beta}] + \frac{3}{4} N,
\end{multline}
where $N$ is the number of particles and $I$ and $J$ are spatial orbital labels.

In figure~\ref{fig:hub_rdm} the relative efficiency is shown for the same Hubbard systems used in figure~\ref{fig:hub_hf}, but using the estimates of $\bra \hat{S}^2 \ket$ from the 2-RDM. Results from these two figures used the same parameters but were taken from different simulations. RDM estimates were calculated every $200$ iterations and a blocking analysis was performed. Once again, substantial improvements are found, with an efficiency increase of over $10^6$ observed for $U/t=0.25$. Some remarkably accurate results are obtained. It is known that the exact value of $\bra \Psi | \hat{S}^2 | \Psi \ket$ should be $0$ for the ground state of this system. For $U/t=0.25$, estimates of $\bra \Psi | \hat{S}^2 | \Psi \ket$ change from $3.1\times10^{-7} \pm 2.5\times10^{-7}$ for the fully stochastic formulation, to $-1.3\times10^{-10} \pm 1.4\times10^{-10}$ for $|D| = 10^5$.

Table~\ref{tab:rdm_efficiency} shows relative efficiencies for molecular systems, for both the variational energy and spin estimators. Once again, significant improvements are seen in all cases, and there is always an improvement with increasing $|D|$ for the range of deterministic space sizes considered here. This suggests using a large deterministic space is sensible, although this has to be weighed against increasing memory requirements.

\subsection{Be$_2$ F12 results}
\label{sec:results_f12}

As a further demonstration of the benefits of semi-stochastic, we consider the calculation of an explicitly correlated correction to the basis set incompleteness error in Be$_2$. The explicitly correlated `F12' approach was first proposed in 1985\cite{Kutzelnigg:TCA68-445,Klopper87}, and has since been refined\cite{Kutzelnigg:JCP94-1985,Klopper:JCP116-6397,Ten-no:CPL398-56,Ten-no:JCP121-117,Valeev:CPL395-190} to become a standard tool to accelerate basis set convergence in quantum chemistry\cite{Kong:CR112-75,Hattig:CR112-4}. The aim of this approach is to complement the traditional wave function expansion (in Slater determinants) by a small set of functions which have an explicit dependence on the inter-electronic distance. These `geminal' functions are crucial for an accurate description of the electronic cusps.

In this work, rather than optimizing the sampled wave function in the presence of the explicitly correlated geminal functions, they are instead coupled after the FCIQMC calculation via an internally contracted multireference perturbative approach ($[2]_{R12}$). This method was first proposed by Torheyden {\em et al}.\cite{Torheyden2009,Kong:JCP135-214105} and first applied to FCIQMC in Ref.~(\onlinecite{Booth2012_2}). This approach allows the calculation of the correction through the sampled one and two-body density matrices (after some rank-reducing approximations). The quality of these corrections will provide a further demonstration of the accuracy of FCIQMC reduced density matrices when using the semi-stochastic approach. It should be noted that in previous applications of this method to FCIQMC, the results were without the semi-stochastic adaptation and approximated the RDMs without the use of a replica sampling (in a biased fashion)\cite{Booth2012_2}. An alternative explicitly correlated approach, where the Hamiltonian is `transcorrelated' via an approximate many-body canonical transformation\cite{Yanai2012} has also been used within the FCIQMC framework.\cite{Sharma2014} While it is still unclear which is the optimal strategy within FCIQMC, in this work we focus on the \emph{a posteriori} approach, in order to demonstrate the accuracy of the sampled RDMs with the semi-stochastic adaptation.

Be$_2$ is a very weakly bound molecule which has resulted in extensive study within the literature, both in theoretical and experimental investigations\cite{Sharma2014, Lesiuk2015, Merritt2009, Patkowski2009, Harkless2006, Tew2006, Gerber2005}. The small binding energy is a stern test for FCIQMC, as it requires careful control over stochastic errors on the order of $\mu E_{\mathrm h}$ for reliable results. These random errors limited the accuracy in the previous FCIQMC study of Ref.~(\onlinecite{Sharma2014}), and as such we expect the improvement due to the semi-stochastic approach to be important.

An all-electron semi-stochastic FCIQMC calculation was performed within the cc-pCVDZ-F12 basis set to calculate 2-RDM estimates. From these 2-RDM estimates, $[2]_{R12}$ corrections were calculated via the MPQC code\cite{Roskop2013}. The binding in Be$_2$ is primarily due to dispersion interactions, and requires very high angular momentum atomic orbitals for an accurate description. As such, it is not to be expected that the cc-pCVDZ-F12 basis set (which only contains up to $d$ angular momentum orbitals) will provide high quality results. Rather, this system is used to demonstrate the very great accuracy with which $[2]_{R12}$ corrections can be calculated when using the semi-stochastic approach.

The FCIQMC calculations used time-reversal symmetrized functions\cite{Smeyers1973, Olsen1988, Booth2011} as basis states, as a compromise between Slater determinants and full configuration state functions. The total space size was $\sim 4 \times 10^{10}$ basis states. The deterministic space consisted of $3 \times 10^4$ states, chosen using our scheme presented in section~\ref{sec:determ_space}. $5$ repeats were performed for each geometry (starting from different random number generator seeds) to provide (uncorrelated) estimates with which to estimate error bars on the [2]$_{R12}$ corrections, and also on $E_{\textrm{RDM}}$ estimates. Approximately $6 \times 10^6$ walkers were used for each calculation. This was found to reduce initiator error to no greater than roughly $10\mu E_h$ at each bond length, with such accuracy being deemed necessary for this weakly-bound system. An optimal value of the single parameter in the F12 geminal (the $\gamma$ exponent) was found by minimizing the $[2]_{R12}$ correction at equilibrium geometry, yielding an optimized $\gamma = 2.44 a_0^{-1}$.

\begin{figure}
\includegraphics{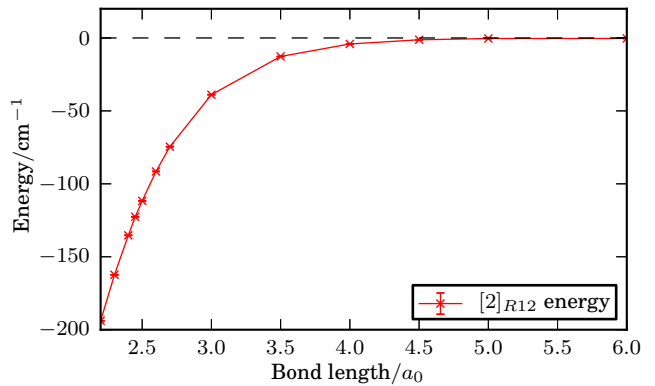}
\caption{The $[2]_{R12}$ energies (relative to the FCI value at infinite separation) for Be$_2$ in a cc-pCVDZ-F12 basis set, with all electrons correlated and $\gamma = 2.44 a_0^{-1}$. Error bars, which are each calculated from 5 independent estimates, are plotted but not visible. All error estimates are less than $0.12$cm$^{-1}$ ($0.5 \mu E_h$). It is seen that the result at large bond lengths approaches the FCI result with very great accuracy.}
\label{fig:be2_r12}
\end{figure}

In figure~\ref{fig:be2_r12} the $[2]_{R12}$ contributions are shown. These values are plotted relative to the correction at infinite separation. This infinite separation value was calculated by using the exact FCI 2-RDM for the Be atom, and therefore has no stochastic or systematic error. It is therefore very encouraging that our FCIQMC results converge to this value so accurately. Error bars are not visible on the plot, all being less than $0.12$cm$^{-1}$ (0.5$ \mu E_h$). This demonstrates that the semi-stochastic adaptation works well for these $[2]_{R12}$ corrections, as for the RDM-based quantities already considered.

\begin{figure}
\includegraphics{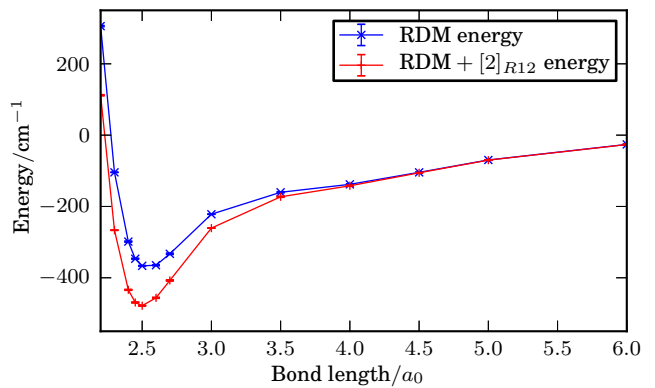}
\caption{Be$_2$ binding curve in a cc-pCVDZ-F12 basis set, calculated by combining $E_{\textrm{RDM}}$ and $[2]_{R12}$ contributions. All $8$ electrons are correlated and $\gamma = 2.44 a_0^{-1}$. The addition of the $[2]_{R12}$ basis set incompleteness correction improves the energy estimate. However, the system is still under bound by around $450$cm$^{-1}$ compared to experimental values, suggesting that larger basis sets (with high angular momentum functions) are still required. The relatively small improvement of the $[2]_{R12}$ correction for this system and basis set is not related to the use of FCIQMC. Error bars, which are each calculated from 5 independent estimates, are plotted but not visible. All errors estimates are less than $1.8$cm$^{-1}$ ($8 \mu E_h$).}
\label{fig:be2_combined}
\end{figure}

In figure~\ref{fig:be2_combined} the variational energy calculated from the 2-RDM estimates (using Eq.~(\ref{eq:rdm_energy})) is shown, together with the total energy, including both $E_{\textrm{RDM}}$ and $[2]_{R12}$ contributions. It is seen that the $[2]_{R12}$ corrections greatly reduce the basis set incompleteness. However, for this system the cc-pCVDZ-F12 basis, despite the addition of the explicitly correlated correction, is still not sufficient to obtain results compatible with the most accurate estimates\cite{Merritt2009, Patkowski2009} of the well depth, which are around $930$cm$^{-1}$.

Interestingly, the $[2]_{R12}$ correction for this system is not as significant as for many previously-studied systems. It has been demonstrated that results of aug-cc-pVQZ quality can sometimes be obtained within an aug-cc-pVDZ basis\cite{Torheyden2009}. The small improvement here is perhaps explained by the F12 corrections being less effective at describing bonding via dispersion interactions. For the Be atom, an accurate variational energy is $-14.66736E_h$\cite{Komasa2002}. By including both $[2]_{R12}$ and $[2]_S$ corrections, the cc-pCVDZ-F12 Be energy goes from $-14.6574E_h$ to $-14.6691E_h$ (with $\gamma=2.44 a_0^{-1}$). Thus, the accuracy of the atom energy is greatly improved, although interestingly it appears that the energy is below the complete basis set limit by $\approx 2mE_h$ (possible due to this perturbative framework). The improvement to the Be$_2$ binding energy is less significant. We emphasize however, that relatively small improvement of $[2]_{R12}$ in this case is not a result of using FCIQMC. In particular, the semi-stochastic adaptation is found to perform extremely well, with all stochastic errors being less than $1.8$cm$^{-1}$ ($8 \mu E_h$).

\section{Conclusion}
\label{sec:conclusion}

We have performed a detailed study of the semi-stochastic adaptation to FCIQMC and presented a new method for generating the deterministic space. This approach creates the space naturally from the dominant states in the FCIQMC wave function and avoids having to perform multiple large and time-consuming deterministic ground-state calculations. Using this approach greatly improves the stochastic efficiency of calculations for a large range of systems and parameters.
We also demonstrated that a simpler approach to the initiator approximation in semi-stochastic FCIQMC gives the same benefits as the approach previously suggested by Petruzielo \emph{et al}\cite{Petruzielo2012}.
We have also explained how the semi-stochastic adaptation can be implemented with relative ease in an existing parallel FCIQMC code. It has been shown that the parallel scaling is not significantly worsened for a range of numbers of CPU cores. Rather, for systems where parallel efficiency is particularly poor, it has been shown that deterministic approaches can significantly improve the parallel performance and speed up FCIQMC calculations. This is a significant result and deserves further investigation.

The benefit of the semi-stochastic approach was demonstrated for FCIQMC estimates of the 2-RDM. This is significant as we expect this object to be important in future applications of FCIQMC, as it can be used for calculating non-trivial two body operators, variational energy estimates and F12 basis set incompleteness corrections. We observed improvements in efficiency to factors in excess of 1 million. As a concrete demonstration, these improvements were applied to the calculation of F12 corrections for Be$_2$.
We therefore suggest that using the semi-stochastic adaptation should become standard practice when performing FCIQMC calculations.

\section{Acknowledgments}

We thank C. J. Umrigar for helpful comments on the original manuscript. We also thank D. P. Tew for helpful discussions.
N.S.B. gratefully acknowledges Trinity College, Cambridge for funding.
J.S.S. acknowledges the research environment provided by the Thomas Young Centre under Grant No.~TYC-101. G.H.B. gratefully acknowledges the Royal Society for a university research fellowship. This work has been supported by the EPSRC under grant no. EP/J003867/1.


\begin{thebibliography}{48}
\bibitem [{fn1()}]{fn1}%
  \BibitemOpen
  \href@noop {} {}\bibinfo {note} {If the deterministic space is the entire FCI
  space, then the algorithm reduces to the power method without explicit
  normalization; FCIQMC can be viewed as a stochastic version of this
  approach.}\BibitemShut {Stop}%
\bibitem [{fn3()}]{fn3}%
  \BibitemOpen
  \href@noop {} {}\bibinfo {note} {We do not include the semi-stochastic
  initialization time because this would lead to the efficiency measure
  depending on the number of iterations performed, which is undesirable. We
  note, however, that this initialization time is usually negligible compared
  to the rest of the FCIQMC simulation, taking at most a few minutes for a
  deterministic space size of $10^5$, in our implementation.}\BibitemShut
  {Stop}%
\end{thebibliography}
\end{document}